# The OverRelational Manifesto.

Evgeniy Grigoriev © 2005-2006
grigoriev.e@mail.ru

**The OverRelational Manifesto (below, ORM) proposes a possible approach to creation of data storage systems of the next generation. The need for such new systems is stipulated by the fact that the capabilities of the existing DBMS are not satisfactory for development of complex information systems. In essence, the matter in question is the possibilities of an adequate description of a complex enterprise.**

**The most important part of the data storage system created on the basis of the approach proposed is an object-oriented translator operating in the environment of a relational DBMS. The expressiveness of such a system is comparable with that of object-oriented programming languages. On the other hand, the approach proposed admits the fundamentality of the relational model of data [2, 3], including the possibility of using that model as a formal basis for a data storage system of the next generation.**

## INTRODUCTION

The possibilities of uniting the properties of object-oriented and relational systems within the framework of a single system have been vividly discussed by specialists for a long time. About ten year ago, in a comparatively short time interval, three papers were published in which different groups of authors declared the set of necessary properties such a system should possess.

"The Object-Oriented Database System Manifesto" [M1] (chronologically, the first one) was created by adherents of object-oriented databases. Making their choice, they actually discard the relational model as a hangover. In their opinion, the object database is just the object-oriented programming environment that realizes the property of long-term storage of objects and augmented with tools for retrieval of necessary information, i.e., by a system for generation and execution of queries.

In contrast, "Third-Generation Data Base System Manifesto" [M2] (the second one) proposes an evolutional approach, according to which the data storage systems of the next generation should grow out of existing systems for data storage inheriting all their features. The authors list the properties, which the new generation databases should (in their opinion) possess. In the list they unite the advantages of object and relational systems and assert that the useful properties should be implemented in a DB- programming language, which should be based on SQL.

The authors of the "The Third Manifesto" [M3] disagree with the first manifest and are of the opinion that the DBMS of the third generation should be based on a mathematically rigorous relational model. They suppose that systems based on this model should be augmented with an extendable system of scalar types(domains). While the currently existing relational systems have a rather poor set of domain, the DBMS of the new generation should have an extendable one. Thus, the attribute of tuple of relation may be both a simple value (for instance, a number or a string of fixed length) and an arbitrary complex <u>scalar</u> value. The authors of the "The Third Manifesto" also disagree with the second manifest since they extremely disapprove the SQL language. They reasonably claim that this language distorts the relational model. Other ideas of the second manifest, in particular, the idea on the correspondence between types and relations are also seriously criticized. The OverRelational Manifesto treats this criticism as justified and righteous.

In contrast with the previous manifests, the "The Third Manifesto" is formal and logical. However, ORM cannot unconditionally accept the claims of the "The Third Manifesto", because, in our opinion, the premises, which are its basis, are incomplete. Recall that, answering the first question "*What concept in the relational world is the counterpart to the concept* "object class" *in the object world?*", the "The Third Manifesto" considers the two possible versions
   1. domain = object class
   2. relation = object class
"The Third Manifesto" argues strongly that the first of these equations is right and the second is wrong (ORM quite agrees with this) and, further, its arguments are based just on the first version.

Note that ORM does not claim that the propositions of the "The Third Manifesto" are erroneous. However, ORM does not doubt that the first answer (even the right one) to the question in the preceding paragraph is not a complete answer to the question about the relationship between the "object world" and the "relational world." There is another approach that can be described by none of the answers proposed in the "The Third Manifesto". Nevertheless, this approach allows us to unite the properties of object and relational systems within the framework of a united system. This approach is considered below.

## Main requirement of ORM

In relational databases all data are represented as a set of relations [3]. We assume the data describing a certain enterprise, which is a set of entities. There is a complex correspondence between the modeled entities and the set of relations that describe these entities, namely, data about any entity may be included in many different relations and any relation may include data about many different entities. Possible versions (for instance, where only one tuple of one relation corresponds to an entity) may be treated only as a particular case of the situation described.

There is only one rule that holds in any case. Note that the enterprise can be treated as a complex entity consisting of many entities. The whole enterprise is described in the relational database as a set of relations. A situation can be considered where the database stores the information about only one of these simpler entities. But anyway the corresponding value stored in the relational database is a set of relations. The matter in question is any enterprise and any its entity - the data describing their state must be represented in the relational database as a set of relations, because this is the main requirement [3] of relational databases.

We assume that the value describing the state of any entity is a set of relations (or, more definitely, relation values), which is a subset of the relational database describing the state of the whole enterprise.

ORM claims that a system that allows to specify <u>explicitly</u> and to manipulate such subsets is the required system that possesses the properties of both object and relational systems. Accordingly, the **main requirement** of ORM is the following:
**The value describing the state of an entity of an enterprise must be represented as a set of relation values.**
Any system satisfying the main requirement will be referred to as an R*O-system.

Remark. Therefore, to relate the "object world" and the "relational world", ORM associates object with a set of relations. Note that the concept of database appearing in RMD is also defined as a set of relation. Essentially, ORM regards the database as a collection of subsets (not necessarily disjoint, even embeddings are possible); by definition, each subset may also be called database.

# Part 1. R*O-system from outside.

The type system necessary for description and manipulation of data, constraints on the data integrity, and a set of operations are described. It is shown that complex structure definition, in which these types are used, can be treated as definition of set of relational variables (R-variables). The common rule for definition and naming of possible R-variables is formulated, which asserts that the definition of complex reference structure, in which path expression $n_1.*^1.*^2.n_z$ is correct, can be interpreted as definition of a relation variable named as $n_1.*^1$, in which the scalar attribute with a name $*^2.n_z$ exists. We also consider the main control commands of the system.

**Types of an R*O-system**

The data in an R*O-system are described as a set of values of a predetermined or a constructed type. The types are partitioned into value ones and object ones. Value types describe values, while object types describe objects.

The only way to distinguish values is to compare them directly and completely. In other words, values identify themselves. This is their principal distinction from objects, which are identified with unique object identifiers (OID) used for distinguishing any object from other ones.

The value types are the following:
1. scalar type including the basic ones (numerical, symbolic, Boolean, etc.) and reference types (they will be described below). A value of the scalar type will be referred to as a scalar.
2. constructed tuple type. A value of this type (hereinafter, a tuple) is a set of pairs "an attribute name, a value of the attribute of the scalar type." Accordingly, the tuple type is defined as a set of pairs "attribute name, scalar type of the attribute."
3. constructed set type. A value of this type (hereinafter, a set) is a set of scalar or tuple values. Accordingly, a set-type variable is defined as (*variable_name* AS SET OF *name_of_scalar_or_tuple_type*).

<u>Remark.</u> We skip the possibility of existence of other methods for defining the scalar type, but we admit that such ways may exist. For instance, a type may be defined by an explicit enumeration of its values. We also skip the possibility of existence of collection-types different from the set-type; however, we admit the existence of such types on the condition that the values of these types may be unambiguously transformed into values of the set-type and, conversely, from these values into the initial values

> ***Example.*** *As an example, consider a simple database for description of movement of goods and their storage on a set of warehouses.* `INTEGER, FLOAT, DATE, STRING` *are basic scalar types. Note that this example is intended exclusively for demonstration of some specific features of R*O-systems and, therefore, makes no pretence of completeness and accuracy.*
>
> *We describe* `ArtQty` *as a tuple type. The* `Art` *attribute is of the scalar reference type* `Article` *(described below).*
> ```
> DESCRIBE TUPLE ArtQty
> {
>    Art Article;
>    Quantity INTEGER;
> }
> ```

A system element intended for storage of scalars will be referred to as a *field*. Accordingly, an element for storing tuples is an unordered collection of fields and an element for storing values of set-type is a set of such collections.

Object types describe objects. An object has a unique identifier (OID), which expresses its property of uniqueness and identifiability and is used for organization of access to this object. The unique object identifier is separated from the values of its components.

***Example.*** *The object type* `Brand` *describes the unique trademarks of goods.*
```
CREATE CLASS Brand
{
  Name STRING
    CONSTRAIN GLOBALKEY Name;
}
```
*(The integrity constraint* `GLOBALKEY` *is considered below)*

*The object type* `Article` *describes types of goods. Objects of this type have a unique field* `No`. *Each type of goods belongs to one of the brands.*
```
CREATE CLASS Article
{
  No STRING
    CONSTRAIN GLOBALKEY No;
  BrandName STRING
    CONSTRAIN FOREIGNKEY BrandName ON Brand.Name;
}
```
*(The integrity constraint* `FOREIGNKEY` *is considered below).*

Object types are constructed types. The definition of an object type consists of a specification and an implementation [6]. A *specification* is a declarative list of external properties (attributes and methods), which may be treated as an interface used for organization of interaction with this object. An *implementation* is an externally inaccessible totality of data structures and programming code, which realizes the specification of this type on the basis of types and operations existing in this system.

We consider the attributes and methods of object types as components, which contain or return some values, i.e., have value type. The specification of a method may be treated as the specification of the component whose name coincides with the name of the method and whose type coincides with the type of the value returned by the method (the specification of a method also includes a description of the parameters of the value types).

Thus, the specification of an object type is the totality of specifications of the value components. The totality of values of the object components determines the state of the object. Any component can be implemented as either a one that stores a value, or a one that calculates the value; however, it is important to understand that the specification of the component does **not** define whether this value is stored or calculated.

Remark. Of course, in a sense, the specification determines the implementation. For instance, if a method has some parameters, then it may be supposed that the value returned by the method is calculated. Nevertheless, this dependence is not obvious. An implementation is quite possible where no calculations are needed for the value returned by the method (for instance, when inheriting, one of the method implementations may use parameters while another one does not use them). Moreover, this value can be implemented as a stored one. On the other hand, an implementation of attributes having no parameters may contain calculating expressions.

For object types, local, global, and foreign keys may be specified. They are defined as a totality of scalar fields of the object. The keys may be simple (which contain only one scalar field) and complex (which contain several scalar fields). The keys are the data integrity constraints.

A local key may be specified for set-components. It contains the fields that specify the uniqueness of scalars or tuples present in the set within the object component (of course, the scalars or the tuples may be repeated in other objects). If the local key is not explicitly defined, then it is meant that the elements of the set are distinguished by their complete values. In this case, the structure of the key coincides with that of the elements of the set.

<u>Remark.</u> In this connection, note that, for the set of scalars where the structure of the key cannot be different from the structure of an element of the set, its explicit definition is meaningless.

A global key explicitly shows that the object is different in its state from other objects of the same type. Global keys are optional, because, in any case, the object identifier determines the fundamental uniqueness of the object. The global key may contain either scalar components, or scalar fields of a tuple-component or of a set-component. Several global keys may be specified for one object type

A global key specified as a totality of fields of a set-component shows that the elements of this set are unique in all objects of this type existing in the system (and within each object). Note that, if, for a type, a global key is specified on fields of a set-component, then, for each object of this type, there may simultaneously exist a set of unique values of this key.

Foreign keys contain either scalar components or scalar fields of one of the tuple components or set-components analogous to the fields present in one of the existing global keys. The meaning and goals of the foreign keys are similar to those of the foreign keys of relational database management systems.

*__Example.__ The object type* `Warehouse` *describes warehouses. No global key is specified for objects of this type. Thus, the system may contain objects of this type, which are undistinguishable by value. The component* `ResourceItems` *contains data about the commodity stored at the warehouse. The value of the component* `ResourceItems` *is specified as a set of tuple values of type* `ArtQty`*; here, it is specified that the* `Art` *attribute is a key.*

```
CREATE CLASS Warehouse
{
   Address STRING;
   ResourceItems SET OF ArtQty
      CONSTRAIN
      LOCALKEY Art;
}
```

*The object type* `GoodsMotion` *describes the movement of the goods. The objects of this type, which exist in the system, are unique with respect to the* `No` *attribute. The components* `FromWarehouse` *and* `ToWarehouse` *may refer to the object of the* `Warehouse` *existing in the system (the values of these fields may be undefined -* `FromWarehouse` *in the case of shipments,* `ToWarehouse` *in the case of sales). The component* `MovedItems` *contains the information about the types of goods and about their quantity.*

```
CREATE CLASS GoodsMotion
{
   No INTEGER
      CONSTRAIN GLOBALKEY No;
   DateOfAction DATE;
   FromWarehouse Warehouse;
   ToWarehouse Warehouse;
   MovedItems SET OF ArtQty
      CONSTRAIN
      LOCALKEY Art;
}
```

The totality of value types considered here allows one to uniquely fulfill the main requirement of a R*O-system. Indeed,

- A value of the set-component may be treated as the relational value whose scheme corresponds to the scheme of elements of this set.
  <u>Remark.</u> Considering a set of values of a scalar type, we assume that the corresponding relation scheme contains only one argument **value** of this type.
- A value of the tuple component may be treated as the relational value specified by the scheme of this tuple with a single tuple.
  <u>Remark.</u> We assume that, for relations that unambiguously has only one tuple, it is not necessary to specify a key. It is meant that the keys, which determine the uniqueness of tuples within the relation and enable us to organize access to some of these tuples, are not needed in the case where the relation has only one tuple by default.
- The totality of scalar components included in one object may be treated as a relational value with a single tuple. Such a totality will be referred to as the *own tuple* of the object. The scheme of the corresponding relation is specified while describing the object type and contains all its own (i.e., noninherited) scalar components.
  <u>Remark.</u> In principle, there is no obstacle for considering each scalar component as a value of a unary and single-tuple relation. The union of these attributes in one own relation significantly simplifies our construction.

The considered totality of value types allows us to describe data structures whose complexity is comparable with that of data structures existing in conventional OO languages. Indeed, the object components may be simple values (scalars), records (tuples), and repetitive groups (sets). The existence of the reference type related to the basic scalar types allows us to describe compound nested structures.

Object types form a hierarchy of inheritance (object types cannot inherit from value types and value types cannot inherit from object types). The inheritance of object types assumes that the specification of the type of a descendant includes the specification of the parent type (or the type of an ancestor). Multiple inheritance is admitted. There exists a predefined dummy object type **Object**, which is a default ancestor for any object type.
<u>Remark.</u> Under inheritance from types with a common basic type, the specification of this type is not repeated (in terms of C++, we can say that the specifications of object types are inherited *virtually*). The system must include a mechanism for resolution of collisions of the component implementations, which are possible in the case of multiple inheritance.

***Example.*** *Let us describe a tuple type* `SaleQty`, *which is intended for description of the quantity of a certain type of goods.*
```
DESCRIBE TUPLE SaleQty
{
   Art Article;
   Quantity INTEGER;
   Price FLOAT;
}
```

*The object type* `Sales` *describes the facts of sales. Since sales may be treated as a special case of shipments from a warehouse, the object type* `Sales` *is a successor of the* `GoodsMotion` *type. The* `SaleItems` *component contains data about the traded commodity. Its value is defined as a set of tuple values of the* `SaleQty` *type, where the attributes* `ArticleNo` *and* `Price` *compose a key (the same commodity may be sold at different prices). The* `IsPayed` *component indicates whether the payment was made. The* `DoSale` *component is a method that takes (as an attribute) the date of shipment and returns the result testifying the success of the shipment operation.*

```
CREATE CLASS Sales EXTENDED GoodsMotion
{
   IsPayed BOOLEAN;
   SalesManager Manager;
   SaleItems SET OF SaleQty
      CONSTRAIN
```

```
      LOCALKEY (Art, Price);
   DoSale (DateOfSale) BOOLEAN;
}
```

*The scheme of data may be changed. For example, suppose that the object type* `Brand` *should be extended with a component containing some information about the sales of commodities that belong to a certain (*`this`*) trademark. Note that, since each commodity belongs to only one trademark, the* `Art` *field is described as a global key. Thus, any tuple of the* `SaledItems` *component of any object of the* `Brand` *type is unique in all these objects.*
```
ALTER CLASS Brand
   ADD SaledItems SET OF ArtQty
      CONSTRAIN
      GLOBALKEY Art;
```

In the general case, the specification of an object type includes
1. the type name;
2. a list of parent types (unless otherwise is defined explicitly, the parent type is the **Object** type implicitly);
3. a collection of specifications of components, which include (a) the component name; (b) the value type of the component, and (c), optionally, the set of parameters each one described as a pair ;
4. a set of data integrity constraints, i.e., keys.

<u>Remark.</u> We do not consider other possibilities of specifications of object types, which are typical of programming languages, (for example, visibility modifiers private and public; updatability modifiers readonly, etc). However, we admit that such possibilities can exist and be useful.

An implementation of a type is the totality of implementations of its components. The implementation of any component defines the source of the value of this component indicating whether it is stored or calculated. In the latter case, it contains a calculating expression or a calculating function. The calculating expressions and functions can take arguments of value types. The component implementation may contain the predicate of the component, i.e., constraints on its possible values determined by the enterprise.

**Example.** *Implement the object type* `GoodsMotion` . *All of its components are stored.*
```
ALTER CLASS GoodsMotion
REALIZE No, Date, FromWarehouse, ToWarehouse, MovedItems AS STORED;
```

*The* `Article` *type is implemented just in the same way*
```
ALTER CLASS Article
REALIZE * AS STORED;
```

When a component is inherited, its implementation may be changed. Thus, the type components (hence, the type itself) are polymorphic in the sense that several implementations may correspond to the same specification.

<u>Remark.</u> Actually, ORM claims that not only methods, but also attributes of objects may and must implement the basic OO concepts such as encapsulation, inheritance, and polymorphism. This claim is based on the fact that the set-theoretical and special operators of relational algebra can be applied to relation values regardless of whether these values are stored or calculated. This allows us to divide the description of the components into specification and implementation. In the specification, the signature of component is defined, namely, its name and its value type. The implementation defines the source of the attribute value, indicating whether it is stored or calculated. In the latter case, it contains the calculating expression. Thus,
- Components of an object are encapsulated. In the public specification of this type, only the object signature is defined. The description of the implementation of the element that contains the information about the source of the component value is hidden.
- The object component are inherited. The specification of the successor-type includes the specification of the basic type including the component specifications defined in this specification.

- Components of object types may be polymorphic. The implementation of types may vary in the inheritance process. This, in particular, implies that a component defined as stored in the parent type may become calculated in the successor type (and vice versa) and that the calculating expression may be changed in the inheritance process. Thus, the mentioned OO concepts may be implemented by all components of objects, namely, by methods, as well as by attributes.

**Operations on objects**

Objects may be created and deleted. To change the state of an object, it is necessary to refer to some components of this object, which are defined in the appropriate object type. The state of an object can be changed both explicitly (explicitly specified operation changing the component value) and inexplicitly (by executing some methods).

The READ operation can be applied to any component. To components without parameters (i.e., attributes), the ASSIGN operation can also be applied. Of course, the value assigned to an attribute should have the type of this component. To set-attributes, we can also apply operations, which change the number of tuples (INSERT, DELETE) and the state of the existing tuples (UPDATE).

<u>Remark.</u> We understand the difficulties that can arise when implementing operations, which explicitly change the attribute value in the case where these attributes are implemented as calculated (note that modern DBMS can solve such problems - for example a trigger can be set on a view).

<u>Remark.</u> Admitting that the ASSIGN operation is allowed for all attributes without exception, ORM does not claim that this operation must *change* the value of this attribute inevitably. This concerns the calculated attributes. For instance, an attribute containing some information about the warehouse balance is calculated on the data about deliveries and shipments. Hence, figuratively speaking, the system must neglect the attempts of assigning some values to this attribute, if the contrary is not specified by the implementation, of course.

<u>Remark.</u> Speaking about the correspondence between the operand types in the assignment operation, we skip the possibilities of implicit type casting (for instance, arithmetical). These possibilities exist in many languages. In our opinion, they are possible and useful.

The methods are sequences of the operations on the components defined in the object type, as well as on variables, which are visible in the body of the method (among them are the global R-variables described below). Methods may have local variables of value types. The lifetime of the local variables is bounded by the execution time of the method.

Object type may have constructors and destructors, i.e., methods invoked when creating or deleting an object, respectively.

***Example.*** *Realize the* `DoSale method of type Sales`
```
ALTER CLASS Sales
   REALIZE DoSale
AS BEGIN
IF DateOfAction NOT IS NULL THEN //if the shipment is made
   IF DateOfAction = DateOfSale Then Return TRUE;
      //and on the same date – OK!
   ELSE RETURN FALSE;
      //the shipment is made on another date - error
ELSE //shipment is not made yet
   IF IsPayed THEN
      BEGIN
         DateOfAction := Date Of Sale;
         RETURN TRUE; //the sale is paid - OK
      END
   ELSE
      RETURN FALSE;//the shipment is impossible (is not paid) - Error
END;
```

For objects, two operations are defined, which allow one to determine the type of the object. Since R*O-systems support the inheritance of object types, this requirement should be explained. There are two operations that allow one to determine the type of an object. The first operation **o IS t** (where **o** is a reference to the object and **t** is the type name) returns true if the object specified by the reference **o** <u>is</u> an object of the specified type. It is assumed that, in the case of inheritance, an object of any successor type is also an object of the parent type. In particular, this implies that the operation **o IS Object** (where **Object** is a predefined dummy type) returns "true" for any object. The second operation **o OF t** returns "true" only if the object specified by the reference **o** <u>was created</u> as an object of class **t** (i.e., this object was created by the **new t** operation). Accordingly, **o OF Object** is always "false".

**OID and references**

Recall that each object existing in the system is identified by its unique object identifier (OID) that the system assigns to the object when creating it and which distinguishes this object from any other object of any object type. The OID contained in the reference variable provides an access to components of referenced object, which are defined for the corresponding object type.

The operation of object comparison (the same object – the different one) is based on direct comparison of their object identifiers. This is why the object identifiers (by themselves) are considered as the values of the reference scalar type (domain) **DOID**. The fields of reference type present in the object components allow one to describe the existing relationships between the objects of the modeled enterprise. The operations of assignment, comparison, and implicit dereferencing are defined for variables of the reference type. The last one means that any operation except for the operations of assignment and comparison is executed on the object (not the reference variable) referenced with this variable.

At any time instant, the set of active values of the reference type in the system is restricted by the set of values of the object identifiers of the objects which exist in the system at this instant. Each object type is associated with its reference type. This reference type is created simultaneously with the object type. The names of these types coincide. The reference types form an inheritance hierarchy similar to the inheritance hierarchy of object types. In the case of reference types, the inheritance means that the reference to an object of a certain type may contain OID of any object of this type, including OID of objects of any successor types. In the system, a reference type **Object** is predefined. A field of this type may refer to any object existing in the system.

The system may contain variables that are sets of references to objects of a given type (variables of set types defined on the reference type or, in other words, group references). Note that the value of a group reference may be treated as a value of a relation. A single reference (i.e. a reference to only one object) can be treated as a particular case of a group reference.

**R-variables**

R*O-system allows organizing group access to objects and data of these objects, which is based on the relational data model. The possibility of such an access is based on the fact that the object type declaration may also be treated as the declaration of a set of relation variables that contains data of all objects of this type existing in the system. They will be called R-variables. Consider these variables in more details.

<u>R-variables of components of the object type.</u>
As has already been said, the state of an object is described by a set of relational values defined on the set of scalar types. Any object gets a unique object identifier, which may be treated as a value of scalar type. Hence, the declaration of the object type **t** containing a component **a** with the scheme $(x_1:D_1, …, x_n:D_n)$ may also be considered as the declaration of a variable **t.a** of the relation with the scheme $(OID: DOID, x_1:D_1, …, x_n:D_n)$ (note that this relation is defined on the same set of scalar types). One can see that the name of this variable is defined as the concatenation of the name of the type and the name of the component of this type (we use the dot notation). It should be noted that the relational model imposes no constraints on the names of the relation variables and relation attributes other than their uniqueness.

The variable **t.a** contains the <u>totality of values</u> of the component **a** of all objects of object type **t** existing in the system in the way, that each tuple of component **a** of any object of type **t** is associated with the object identifier of this object. This identifier is contained in the OID attribute. Thus, this attribute is the *back reference* to this object.

Note that, characterizing the content of variable **t.a** as "the totality of values," we do not mean that it is obtained as a result of the simple union of these values. For example, being unique in each object, tuples of a set-component **a** of objects of class **t** may be repeated in different objects of this class. Thus, the simple union $v_1$ UNION $v_2$ UNION …, where **v** is the value of a set-component **a** of an object may result in the loss of data that consists in the loss of such repeated tuples. The **OID** attribute of the variable **t.a** uniquely identifies each object

$(OID_1 \times v_1)$ UNION $(OID_2 \times v_2)$ UNION … (where $\times$ is the symbol of Cartesian product)

and guarantees the absence of such a loss of data in that way.

This data representation implies the following. The conventional data access in traditional OO-systems begins with OID stored in a way (i.e., with a pointer to the object). Using this OID, one can obtain access to attributes and methods of the objects (if OID is not stored, the object is assumed to be lost). The variable **t.a** allows one to perform the converse action, namely, to obtain the OID (i.e., the pointer to the object) on the basis of data of the component **a** of objects of type **t**. In particular, this implies that it is not necessary to store the OID of an object in extra special-defined variable to obtain access to this object..

The variables, which are similar to the described variable **t.a**, will be referred to as <u>component R-variables</u> of the object type. A set of such R-variables may correspond to each object type **t**, namely, one R-variable for each component of the tuple type or set type and one more component for the own tuple of the object. In other words, the number of R-variables of a component of any object type is the same as the number of relations describing the state of any object of this type.

<u>Remark.</u> It is important to understand that the values stored in R-variables are always relation values. The expression "R-variable … of type **t**" means only that this R-variable is associated with type **t**.

Note that the value of the OID attribute existing in R-variables is a system value. Because of this, for the access to the OID attribute, we use below the functional expression **Object(Rvar)** instead of expression **Rvar.OID**. Here, **Rvar** is the name of a variable that contains the OID attribute. The expression **Object(expr)**, where **expr** is an expression calculating the relation value with attribute OID, will be used instead of the projection operation **expr[OID]**. In both cases, the expression **Object(…)** returns a group reference. The group reference variable will be considered below as a variable of unary relation with single attribute **Object (ref)**, where **ref** - a name of this reference variable.

<u>Remark.</u> Moreover, obviously, the value of OID itself is of no interest for the user (this value is generated by the system and depends on the realization). Accordingly, an admissible representation of values of the reference type may be completely independent of these values. For example, any value of the reference type may be represented for the user by the string "Object." We may assume that admissible representations can be changed while inheriting. For example, for documents with a number defined as a global key, the admissible representation may be implemented as the string "Document number "…".

*Example. The definition of the object type* `GoodsMotion` *may also be considered as the declaration of an R-variable of a component of type* `GoodsMotion.MovedItems` *with the scheme (OID, Article, Quantity). Accordingly, the projection* `Object(GoodsMotion.MovedItems WHERE Article = "art1")` *returns the value which is a set of OID of objects describing the movement of commodities with article "art1."*

As has already been said, the scheme of R-variable of a component of the type **t.a** is the scheme of component **a** extended with attribute OID. The keys of the variable **t.a** are also uniquely determined by the keys specified for this component. Three cases are possible, namely,
- if a global key is defined for the component **a**, then the key corresponding to the variable **t.a** contains just the same fields as this global key (this is also valid for foreign keys);
- if a local key is defined for the component **a**, then the key corresponding to the variable **t.a** contains the fields present in this local key and the field OID;
- if no key is defined for the component **a**, then the key corresponding to the variable **t.a** contains the only field OID.

R-variable of a type.
Repeat once again that, in accordance with the main requirement of the R*O-system, the state of an object is described by the totality of values of components. These values are values of relations defined on the set of scalar types. The Cartesian product of these values is a value of the 1NF relation that completely describes the state of this object (the scheme of this relation includes all *scalar* fields of all components of this object). With each object, hence, with each such value, a unique object identifier is associated. This identifier is also a value of the scalar type. This allows us to consider the declaration of the object type **t**, as well as the declaration of the variable **t** that contains data of all objects of this type existing in the system (this relation is also defined on the set of scalar types. Like component R-variable, this variable contains attribute OID which is back reference attribute.

Note that the operation, which yields the value of the 1NF relation completely describing the state of an object, must be more complicated than the ordinary Cartesian product. The matter in question is that the fields presented in different components of an object may have the same name. To avoid these collisions of names, we suggest extending the field names with the component names while constructing the Cartesian product. For example, tuple types $R_1$ and $R_2$, on which components $a_1$ and $a_2$, are defined, may have fields with the same name **x**. (Note once again that the relational model imposes no restrictions on the names except for their uniqueness.) In the dot notation, the extended names take the form $a_1.x$ and $a_2.x$. In our opinion, such an action (let us call it "name refining") preserves the semantics and allows expressing the complexity of the object structure by the complex name of an attribute of an R-variable. For example, the set of values of an attribute **x** of a component **a** of a type **t** may be obtained by the selection operation either from the R-variable of component **t.a[x]**, or from the R-variable of type **t[a.x]**.

Thus, a value of variable **t** is a totality of values of Cartesian products of semantically refined components of objects of type **t** existing in the system. Only one variable **t** corresponds to each object type **t**. Such variables will be called R-variables of types.

*Example. The object type* `GoodsMotion` *is associated with an R-variable of this type* `GoodsMotion` *with the scheme (Object(GoodMotion):DOID, No:INTEGER, DateOfAction:DATE, FromWarehouse:Warehouse, ToWarehouse:Warehouse,*

*MovedItems.Article:STRING, MovedItems.Quantity:INTEGER). Note that this relation is defined on the set of scalar types including the type of object identifiers DOID and the reference type Warehouse, which was defined while declaring the corresponding object type. The operation* `Object(GoodsMotion WHERE MovedItems.Article = "art1" AND DateOfAction = '31.05.2005')` *returns a value which is the set of OID of the objects describing the movement of commodities with article "art1" produced on May 31, 2005.*

**Properties of R-variables**

We assume that the data stored at these variables are always actual, i.e., any change in the states of the existing objects implies the unconditional change in the values of the corresponding R-variables. In other words, the data represented in the form of values of object are the *same* data as those represented in the form of values of R-variables (in what follows, we speak about the <u>dual</u> data representation).

Considering the properties of R-variables related to the type **t** (i.e., variables of type **t** and variables of components of this type **t.a**), we should note that these variables exist independently of the existence of objects of this object type and of the number of these objects. The existence of these variables is determined at the instant of type creation. Thus, these variables are global and may be used as operands of nonprocedural commands, as well as in any system-defined procedures and functions.

This approach makes the user free of the necessity of performing any actions directed to the organization of the group access to data (we mean "extends" or special collections existing in some OODBMS). The declaration of type **t** is at the same time the declaration of the corresponding R-variables in which the data are always actual. In this case, in order to denote R-variables, the names are used which are introduced with type description. Such names will be referred to as <u>multi-meaning.</u> The necessary interpretation of multi-meaning names is determined by the operation, which uses these names. For example, in the operation of creation of a new object **new t**, the name **t** is interpreted as a type name. However, in operations of the group data access, the name **t** should be interpreted as the name of an R-variable.

Remark. This approach excludes the ambiguities that are typical of the term "class" which describes something that is at the same time an object factory (~type) and an object store (~variable). We do not use this term in our presentation.

***Example.*** *Le us realize the object type* `Warehouse`. *It contains both stored and calculated components. In particular, the component* `ResourceItems` *is calculated as the difference between the quantities delivered to a certain (*`this`*) warehouse and shipped from this warehouse. Note that, in the calculating expression, we use a global R-variable of type* `GoodsMotion`.

```
ALTER CLASS Warehouse
REALIZE Address As STORED
REALIZE ResourceItems AS
SUMMARIZE (
   SUMMARIZE
   (GoodsMotion WHERE ToWarehouse = this)
   BY Art ADD Sum(MovedItems.Pieces) AS SumPieces
   UNION
   SUMMARIZE (GoodsMotion WHERE ToWarehouse = this)
   BY Art ADD Sum(0-MovedItems.Pieces) AS SumPieces)
BY Art ADD Sum(SumPieces) AS Pieces;
```

As has already been said, R-variables allow us to obtain the reference to an object of type **t** using data of components of objects. In particular, it means that it is not necessary to store OID in a extra special-created reference variable to obtain access to this object. So, after creating an object, we do not need to get OID to store. Hence, we can use the `new` operator, which creates the object, as a nonprocedural command.

<u>Remark.</u> In particular, this implies that it becomes especially important to use constructors that allow us to initialize the object components while the object is created (an object created by the new command without use of the constructor contains no data which make it possible to distinguish it from other objects of the same class).

*Example.* *Suppose that, for the Article type, a constructor is created, which takes, as a parameter, the article name.*
```
ALTER CLASS Article
   ADD Article( InArticle As STRING);

ALTER CLASS Article
   REALIZE Article AS
   BEGIN
      No := InArticle;
   END;
```

*With the help of the command*
```
new Article("art1");
```
*we create a new object of the class* `Article`. *Note that, in this case, we did not store any reference to this object. However, it can always be obtained by selecting from the R-variable of the type* `Article`. *For example, the local variable* `refArt` *of the reference type* `Article` … *declared in the method body*
```
BEGIN
…
refArt Article;
refArt := Object(Article WHERE No = "art1")
…
END
```
*…is initialized by the OID to the before-created object. The integrity constraint GLOBALKEY, which is set the field* `No`, *guarantees that the* `WHERE` *operation returns a reference to only one object.*

**Inheritance and polymorphism in R*O-system.**

As has been said, the specification of an object type does not define whether the values of its components are stored or calculated. A component implemented in the parent type as a stored one may be redefined in a successor class as calculated (and vice versa). Accordingly, when dealing with a polymorphic inherited object type, anyone of R-variables of its components can simultaneously contain both the stored values and the values calculated in different ways.

Strongly speaking, the value of an R-variable of a component should, of course, be calculated and this value is the union **UNION** of several values some of which may be implemented as stored, while the others are calculated. For the R-variables of types, the situation is more complicated. Since the value of an R-variable of type is defined as the Cartesian product of the component values, it can occur that, in some tuples, only several attributes are stored. Because of possible changes of the implementations of components containing these attributes while type inheriting, these attributes may be changed in other tuples as calculated (i.e., figuratively speaking, the R-variable of type is stored *in a tiled form*). However, in any case, the system must calculate the values of any R-variables <u>implicitly for the user</u> (on the basis of information about the type inheritance and implementation of components of these types). To use R-variables, only the type specification is necessary.

The following analogy may be made. Selection constructions are bloated and sensitive to changes in programs. Polymorphic OO-programming languages make unnecessary such constructions, which are required for executing close (in sense) operations on structures, which store data close in their sense. For instance,
```
if s.f=1 then function1(s)
elseif s.f =2 then function2(s)
```
is replaced with invoking the polymorphic method `s.function()`.

Just in the same way, the polymorphic components make unnecessary the explicit use of the **UNION** operator for obtaining a relation value uniting data close in their sense in the case where these data are stored and/or calculated in different ways. Here, it is assumed that the object type is inheritable and that the polymorphic component can change its realization. Thus, if we have an object type **t**, in which a component **a** is defined, and R-variables corresponding to this type are used in queries and methods, then, after creating a type **t\*** that inherits a type **t** and redefines the implementation of component **a**, we have to do <u>nothing</u> to use re-implemented component in these existing queries and methods.

> *Example. Let us realize the object type* `Sales`. *Its component* `SaleQty` *is stored*
> ```
> ALTER CLASS Sales
>    REALIZE SaleItems AS STORED;
> ```
>
> *Realize the component* `MovedItems`, *which describes the commodity shipped from the warehouse. Since its value is directly connected with the quantity of the sold commodity (is equal to this quantity), this component should be redefined as calculated.*
>
> ```
> ALTER CLASS Sales
>    REALIZE MovedItems AS
>    SUMMARIZE SaleItems
>    BY Art ADD Sum(Pieces) AS Pieces;
> ```
>
> *Now, the quantity of the goods stored in the warehouse will be calculated on the basis of data about both shipped goods (`GoodsMotion.MovedItems`) and sold goods (`Sales.SaleItems`). This is achieved at the expenses of modification of the realization of the polymorphic component `MovedItems` in the inheritance process. Note that the schemes and calculating expressions defined earlier are not changed. For example, the expression*
>
> ```
> SELECT SUM(pieces) FROM Warehouse.ResorceItems;
> ```
>
> *which returns the total number of pieces in all warehouses, will return correct values regardless of the existence of successor types for the types `Warehouse` and `GoodsMotion` (of course, under the condition that the components of these types participating in the calculations are correctly predefined). The expression should not be changed even if, in the process of the system operating, such a successor type becomes necessary.*

**Operations on R-variables**

Thus, R-variables give us possibilities of obtaining the reference to an object (or references to objects) by using the component values and, in this way, they allow us to organize the associative access to objects of a given type. The values of these variables are relation values that can be accessed with the help of operations conventional for the existing relational databases. The fact that the structure of these variables is determined by the structure of the corresponding object types makes it possible to introduce two operations with object semantics.

<u>Remark.</u> These operations do not go beyond the framework of the relational model since they can be represented as a superposition of basic relational operations. In particular, this implies that their result is a relation value.

Retrieval objects by values (OV-retrieval operation).
Note that the selection operator **WHERE** seems quite insufficient if the goal of selection is the selection of complex objects which meet certain conditions. This follows from the fact that the condition part of the **WHERE** operator applies to tuples; however, data of each object may be represented in the R-variable by a *set* of tuples. For example, suppose that we deal with objects of a type **t**, in the component **a** of which, the value of relation **R(…, x, …)** is stored, and it is necessary to find objects (i.e., to obtain references to the objects) whose component **a** contains at least one tuple with the attribute value **x = 1** and at least one tuple with the value **x = 2**. Since the attribute **x** in the same tuple cannot be simultaneously equal to one and two, the expression **Object(t.a WHERE x = 1 AND x =2)** is meaningless.

To solve such problems, we propose a operation of retrieval of object by values (OV-retrieval operation). This operation has the form **Rvar<$cond_1$, $cond_2$,…>,** where **Rvar** is an expression determining the R-variable (this can be, for instance, the name of an object type **t** or the name of a reference to this object type **t**) and each **$cond_i$** is a condition applied in the **WHERE** operation.

First, the expressions **<$cond_1$, $cond_2$,…>** are evaluated as **(Object (t WHERE $cond_1$) INTERSEPT Object (t WHERE $cond_2$) INTERSEPT … ).** Hence, the result is a group reference to the objects of type **t** in the system, which satisfies all given conditions. Next, the selection operation on R-variable **Rvar** by the attribute containing OID has to be completed. For example, the expression **t<cond1, …>** is evaluated as **t WHERE EXIST Object(t) $JOIN_{OID}$ (Object (t WHERE $cond_1$) INTERSEPT … )** and represents a subset of the value of the R-variable of type **t**, which contains he information only about the objects satisfying all conditions listed above.

Reference expansion operation.
Another consequence of the fact that R-variables contain back references is possibility to use the group operations for manipulating with references being used to create complex, embedded structures. This is based on the fact that, if objects of type **t** contain a field $x_{ref}$ referring to objects of type **t***, then the R-variable of type **t** may be joined with an R-variable of type **t*** such that the values of attributes **t.$x_{ref}$** and **t*.OID** are equivalent. We will call this operation the reference expansion.

Suppose that we have defined an object type **t** containing a component **$a_n$** defined as a set type **SET OF R**. In turn, the tuple type **R** contains an attribute $x_{ref}$, which is a reference to an object of type **t***. Thus, the object type **t** is connected by reference with the object type **t***. The operation of reference expansion has the form **t EXPAND(a.$x_{ref}$)**. Applying it to the relation of types **t(Object(t), $a_1.x_1$, … , $a_n.x_{ref}$, … , $a_z.x_n$)** and **t*(Object(t*), $a*_1.x_1*$, …, $a*_z.x_m*$)**, we obtain a relation with the scheme **(Object(t), $a_1.x_1$, …, $a_n.x_i.a*_1.x_1*$, …, $a_n.x_i.a*_z.x_m*$, … , $a_z.x_n$)**.

The operation of referenced expansion **t EXPAND(a.$x_{ref}$)** is evaluated as

**t JOIN $_{an.xref}$ (t* RENAME Object(t*), $a*.x_1*$, … , $a*.x_n*$ AS $a.x_{ref}$, $a.x_i.a*x_1*$, … , $a.x_i.a*x_n*$),**

where **a*.x*** are attributes of the R-variable of the object type **t*.**

It is seen that the operation of reference expansion is more complicated than the simple JOIN. The matter in question is that the types **t** and **t*** may have components and component attributes with the same names. Moreover, it is possible that the type **t** and **t*** are one and the same type

(for instance, objects describing people contain references to objects describing parents who are also people). To avoid such collisions, the operation of reference expansion *refines* the component names of the type, to which the reference field refers, using the name of this reference field.

For example, if an object type **t** contains a component **a** with an attribute $x_{ref}$, which refers to an object containing a component **a\*** with an attribute **x\***, then the operation of reference expansion using the dot notation refines the name of the last attribute as **a.$x_{ref}$.a\*.x\***. Recall once again that the relational model imposes no constrains on the names except for their uniqueness. In this case, the complex name **a.$x_{ref}$.a\*.x\*** is surely unique. It is important that this complex name is correct path expression of defined embedded structure. In such a way the operation of reference expansion expresses the complexity of the structure of objects and relations between them in the complex name of the attribute of the R-variable and, thus, preserves the data semantics.

The operation of reference expansion allows us to organize the associative access to data of objects of any type connected with objects of a given type by reference. In this case, the data access is possible both by the reference and in the opposite direction (of course, the actually used relation, which is the result of the **EXPAND** operation, does not mean any direction, because the back-referenced fields containing OID of objects, the reference fields containing OID of related objects, and data fields in them are completely *equivalent*). For example, we can obtain references to objects of type **t** connected by reference $x_{ref}$ with objects of type **t\*** whose attribute **x\*** of component **a** is equal to a certain value, e.g.,
**Object((t EXPAND a.$x_{ref}$)WHERE a.$x_{ref}$.a\*.x\* =1)**

In this case, the same result can also be obtained with using the OV-retrieval operation that is applied to the reference attribute **a.$x_{ref}$** .
**Object(t WHERE a.$x_{ref}$< a\*.x\* =1>)**

Remark.  In this case, the same result can also be obtained with using the OV-retrieval operation that is applied to the reference attribute **a.$x_{ref}$** .
**Object(t WHERE a.$x_{ref}$< a\*.x\* =1>)**
These versions are equisignificant neither in their meaning, nor in the method for evaluating. In the first case, we construct a relation of type **t** expanded by the reference **a.$x_{ref}$** and, then, from tuples with attribute **a.$x_{ref}$.a\*.x\***, we select the OID of objects.
**Object(**
**(t JOIN $_{an.xref}$ (t\* RENAME Object(t\*), a\*.$x_1$\*, … , a\*.$x_n$\* AS a.$x_{ref}$, a.$x_i$.a\*$x_1$\*, … , a.$x_i$.a\*$x_n$\*))**
**WHERE a.$x_{ref}$.a\*.x\* =1 )**
In the second case, first, we select, in the type relation, the tuples of objects satisfying the desired conditions and, then, we use this result when constructing the relation of type **t** expanded by reference **a.$x_{ref}$**.
**Object(**
**t.a WHERE EXIST $x_{ref}$ JOIN ( Object(t\* WHERE a\*.x\* = 1))**
**)**
In our opinion, the second version is more powerful. For instance, it is difficult to find an analog in the style of the first version to the following expression in the style of the second version:
**Object (t WHERE a.$x_{ref}$< a\*.x\*=1, a\*.x\*=2 >))**

The operation of reference expansion can be applied to embedded references
**(t EXPAND a.$x_{ref}$)EXPAND a.$x_{ref}$ .a\*.$x_{ref}$\*[a.$x_{ref}$ .a\*.$x_{ref}$\*. a\*\*.$x_{ref}$\*\*]**

Note that the existence of attributes with refined names means executing operations of reference expansion. So, in cases when expression contains refined names this operation can be implicit. For example, the previous expression can be written simple as
**t [a.$x_{ref}$ .a\*.$x_{ref}$\*.a\*\*.x\*\*]**
Here, the existence of complex refined name **a.$x_{ref}$.a\*.x\*.a\*\*.x\*\*** means that reference attribute **a.$x_{ref}$** and, then, reference attribute **a.$x_{ref}$.a\*.$x_{ref}$\*** have to be expanded in R-variable **t**.

The operation of reference expansion can be applied both to R-variable of type **t** and to R-variable of the component **t.a**, where the reference exists. The result of select operation **t[a.x_ref.a\*.x \*]** is equal to result of select operation **t.a[x_ref.a\*.x \*]**.

**R-variables and references**

Consider the above-introduced group operation **Object(x)** returning OID of all objects, data about which are contained in the R-variable **x**. For example, the statement **Object(t)** returns OID of all objects of class **t** existing in the system.

*Example. Suppose that there is a group variable* `someSales` *of the reference type* `Sales`*. After the operation*
`someSales := Object(Sales WHERE IsPayed = TRUE);`
*this variable contains the set of references to objects of type* `Sales`, *which describe the paid sales.*

Recall that the references are dereferenced, i.e., any operation except for the assignment and comparison operations is performed on the object this reference refers to, but not on the reference itself. This remark also remains true for the operation of access to an object component. If there is a reference **o** to an object of class **t**, then the expression **o.a** describes the access to the component **a** of the object (or objects), to which the reference **o** refers. Accordingly, the expression **ref_t.a** can be treated as the name of an R-variable whose value is the totality of values of the component **a** of the objects of type **t**, references to which are contained in the variable **ref_t**. The value of R-variable **ref_t.a** is calculated as **t.a JOIN ref_t**.

Emphasize the direct analogy between the name of R-variable **t.a** that contains the totality of values of components **a** of all objects of type **t** existing in the system and the name of R-variable **ref_t.a** describing the similar totality for the group of objects specified by the reference **ref_t** (this group may consist of only one object). The schemes of these variables completely coincide. It is assumed that, in both cases, the matter in question is the totality of values of the component **a** of a certain group of objects. This analogy means that these variables may be used in the same operations.

*Example. After the operation*
`someSales := Object(someSales.SaleItems WHERE Price > 100);`
*the above-described variable* `someSales` *of the reference type* `Sales` *contains references to objects of type* `Sales` *describing the paid off sales with the dates of sales whose prices are greater than 100.*

The same analogy is present immediately between the name of type **t** and the name of reference **ref_t** to objects of this type. Similar to the type name, the reference name may be used as the name of an R-variable containing the complete information (in 1NF) about the objects of type **t** to which the reference **ref_t** refers. The schemes of these variables are exactly the same and the value of the R-variable **ref_t** is evaluated as **t.a JOIN ref_t**. Accordingly, the same operations are applicable to these variables.

*Example. After the operation*
`someSales := Object(someSales WHERE DateOfAction = #01.04.2005#);`
*the variable* `someSales` *contains references to objects of type* `Sales` *describing the paid off sales with the dates of sales whose prices are greater than 100 and the sales are dated by April 1, 2005.*

Thus, the reference name (just as the object type name) is a multi-meaning name that must be interpreted depending on the operation, in which this name is used. The reference name (just as the object type name) can be interpreted as the name of R-variable contained value, which is selection of R-variable of type (below referenced R-variable) or component R-variable (below R-variable of referenced component). In essence, the only difference between the R-variable **t** and the R-variable **ref$_t$** is that the first variable is global, while the second is defined only where the corresponding reference variable is defined.

**The rule for defining existence of and for naming R-variables and their attributes.**

Let's return to example where object type **t** with component **a**, which contains the reference field **x$_{ref}$**, was defined (see the description of operation of reference expansion). The result of selection **t[a.x$_{ref}$]** from a R-variable of type **t** is equal to result of selection **t.a[x$_{ref}$]** from a R-variable of a component R-variable **t.a** of this type. This result is the set of OID of objects of type **t\***. Actually, existence of this set is predetermined by path expression **t.a.xref**. Thus, this expression can be considered as a name of the predetermined reference variable containing value of the unary relation, which calculates with any of the stated ways

Similarly to any names of reference variables, such predetermined name can be interpreted as a name of the predetermined R-variable (see the section above). Thus, the definition of objective type **t**, which contains a component **a** with attribute **x$_{ref}$**, which refers to objects of a class **t\*** that contains component **a\*** with attribute **x\*,** can be considered as definition of the following R-variables:
Variable **t** containing, among other, the scalar attribute **a.x$_{ref}$.a\*.x\***, is the result of reference expansion af attribute **a.x$_{ref}$** of R-variable ot type **t**)
Variable **t.a** containing, among other, the scalar attribute **x$_{ref}$.a\*.x\***, is the result of reference expansion af attribute **x$_{ref}$** of component R-variable **t.a**)
Variable **t.a.x$_{ref}$** containing, among other, the scalar attribute **a\*.x\***, is the reference R-variable which is in line with reference **t.a.x$_{ref}$**
Variable **t.a.x$_{ref}$.a\*** containing, among other, the scalar attribute **x\***, is the R-variable of referenced component.

Such reasoning is applicable to structures with any numbers of the embedded references. We shall remind, that object types forming these structures, meet to main requirement of ORM. Operation of reference expansion allows to consider any correct path expression **n$_1$.\*.n$_n$** (where **n$_i$** - anyone, not obligatory different, names, a sign **\*** means any, possible empty, sequence of such names, and **n$_n$** is reference attribute) as a name of the predetermined reference variable for any number of the embedded references. The corresponding R-variable **n$_1$.\*.n$_n$** can contain reference attribute for which the reference expansion operation can be applied also.

Thus, it is possible to assert, that the definition of complex reference structure, in which path expression **n$_1$.\*$^1$.\*$^2$.n$_z$** is correct, can be interpreted as definition of a relation variable named as **n$_1$.\*$^1$**, in which the scalar attribute with a name **\*$^2$.n$_z$** is determined. This rule is the universal rule determining existence and names of R-variable and their attributes in R\*O system.

**Global variables of value types**

Since, in nonprocedural commands of data access, only global variable names may be used, ORM considers as useful and necessary the possibility of defining and using global variables of value types. These variables may be implemented both as stored and as calculated.

*Example. The variable* `someSales` *used in the example in the previous section may be defined as a global variable by the following nonprocedural instruction*

```
CREATE someSales AS SET OF Sales
      REALIZE AS STORED;
```

*Then, the name* `someSales` *may be used in nonprocedural instructions of group data access.*
*We describe a tuple containing a reference to an object of type* `Article` *and a reference to an object of type* `Warehouse`

```
DESCRIBE TUPLE Art2Ware
{
  Art Article;
  Ware Warehouse;
}
```

*Create a global variable containing a set of such tuples*

```
CREATE ArticleOnWarehouse AS SET OF Art2Ware
      CONSTRAIN Art AS GLOBAL KEY
      REALIZE AS STORED;
```

*This global variable may contain, for instance, the information about the distribution of articles over warehouses. The key of this variable (containing the only attribute* `Art`*) shows that any article may be stored on a unique warehouse. Thus, the variable* `ArticleOnWarehouse` *contains the information about a relation of type one-to-many between the set of articles and the set of warehouses.*

<u>Remark.</u> Stored and calculated global variables of value types may be used to emulate tables and views existing in modern relational DBMS.

**Group method invokes**

The group operations on objects somehow or other are operations on components of these objects. Here, by components, we mean both attributes containing the values and methods which return the values (speaking about attributes containing values, we mean exclusively the specification - any attribute can be implemented both as stored and as calculated). In particular, this implies the possibility of the group method invoke

***Expr*.f(…)**, where *expr* is an expression specifying a group of objects, for instance, a type name or a reference name.

Of course, admissible is the superposition of
- conventional relational operations;
- operation of retrieval of objects by values (OV-retrieval) and the reference expansion applied to R-variables of polymorphic types; and
- group invoke of methods (which are also polymorphic).

For example, the expression
**((t WHERE a.x$_{ref}$< a*.x* =1>).method())[x]**
obtains the attribute projection from the result of the group method invoke for objects of type **t** which refer to objects satisfying certain conditions. Note that, despite the explicit object semantics, this expression is fully relational.

***Example.*** *Realize the component* `SaledItems` *of type* `Brand`*, containing the information about sales of articles of a given (*`this`*) trademark. Using the operation of reference expansion* `SaleItems.Art`*, we obtain access to the attribute* `brandname` *of objects of type* `Article`*, which are referred to by objects describing sales.*

```
ALTER CLASS Brand
  REALIZE SaledItems AS
    SUMMARIZE (Sales WHERE SaleItems.Art<brandname = this.name>)
    BY Art, ADD Sum(Pieces) AS Pieces;
```

The following code represents the transaction shipping all non-shipped sales. If this is possible for all sales, then the changes made are accepted. If some sales cannot be shipped, the transaction is rolled back.

```
BEGIN TRANSACTIOM
IF EXIST
   ((SALES <DateOfAction IS NULL>.DoSale(GetTodayDate()) = FALSE)
THEN ROLLBACK
ELSE COMMIT
```

**Brief review of the possible commands of the R*O-system.**

Data types are defined in the data description language (DDL). The types are partitioned into valuable and object. There is a set of basic scalar types. Among instructions for manipulation of value types, the instruction should be distinguished, which allows us to define a new tuple type.

```
DESCRIBE TUPLE tuplename.
{
      scalar_attribute_definition;
      [scalar_attribute_definition;]
}
```

where `scalar_attribute_definition` is an expression describing an attribute of a scalar type. There should exist instructions for manipulation with tuple types and instructions that allow one to delete these types.

A component or a variable of the set type is defined with the help of the SET OF constructor and specification of the used scalar or tuple type.

```
rvarname SET OF tuplename CONSTRAIN [local_keys_definition];
```

where `local_keys_definition` enumerates the fields included in the optional local key.

In the command creating a new object type (and the corresponding reference type), it is necessary to specify this type name, to enumerate the basic type and specification of components, and to define the keys.

```
CREATE CLASS otypename [EXTENDED parenttypename[,parenttypename] ]
{
  value_signature  [CONSTRAIN keys_definition];
  [value_signature [CONSTRAIN keys_definition];]
} [CONSTRAIN keys_definition]
```

Here, `value_signature` is an expression describing a component of value type. The expression `keys_definition` defines the key type and enumerates the fields included in the key.

The operation of altering the type definition may add, change, and delete specifications of proper attributes and methods of the type (i.e., change the specification)

```
ALTER CLASS otypename
      ADD| DROP|ALTER value_signature[CONSTRAIN keys_definition];
```

as well as change the realization of proper and inherited attributes and methods of the type (in accordance with the existing specification).

```
ALTER CLASS otypename REALIZE value_signature AS realize_expr;
```

where `realize_expr` determines the realization of the component, whether it is stored (`AS STORED`), or calculated. In the last case, a calculating expression (`AS valueexpr`) or a calculating function (`AS BEGIN …END`) should be defined for the component.

The operation of deletion of an object type

```
DROP otypename
```

executes actions converse to the actions executed during the addition and deletes records about the realization of attributes and methods of this type.

Nonprocedural commands determining the realization of components and flip-flops contain calculating expressions and procedural expressions describing the actions performed by the flip-flops and methods (in turn, these procedural expressions may call nonprocedural commands of the system control). Thus, the procedural extensions of the control language may be treated as an important part of the data description language (DDL).

The sublanguage for data manipulation (DML) must include
1. commands for creation and deletion of objects of a specified object type.
    - `NEW objecttype(constructor_parameters)`
    - `DESTROY objectgroup`
2. commands for change of the value of the updated attributes of a group of objects of a specified object type
    - assignment operation
    - `INSERT … INTO objectgroup.a`
    - `UPDATE objectgroup.a`
    - `DELETE FROM objectgroup.a`
3. commands for invoking a method for a group of objects of a specified object type
    `EXECUTE objectgroup.methodname(parameters)`
4. expressions based on the known operations of relational algebra extended with the OV-retrieval operation and the operation of reference expansion.

The group of objects `objectgroup`, on which one or another action can be performed, may be defined by explicit indication of data contained in these objects `t< cond1,…>`, references, and other expressions. It may occur that the group contains only one object (for instance, in the case of selection by field being a global key):
`EXECUTE t<GlobalKeyField=1>.method(…)`

Control commands must include commands for creation and deletion of global variables of value types
- `CREATE value_signature[CONSTRAIN keys_definition]`
- `DROP global_value_name`

and for manipulations with values of these variables
- `INSERT … INTO global_value_name`
- `UPDATE global_value_name`
- `DELETE FROM global_value_name`

Access to data in these variables must be performed by expressions based on known operations of relational algebra.

Concluding the first part, emphasize its key points. ORM starts from the known requirement that any information in a relational database is represented by a set of relation values. Accordingly, it is assumed that the information about any entity of an enterprise must also be represented as a set of relation values (the main requirement of ORM).

A system of types is introduced, which allows one to fulfill the main requirement. The data are represented in the form of complex objects, and the state of any object is described as a set of relation values. Emphasize that the types describing the objects are encapsulated, inherited, and polymorphic. Then, it is shown that the data represented n the form of a set of such objects may also be represented as a set of relational values defined on the set of scalar domains (dual data representation). In the general case, any object type is associated with a set of relation variables (R-variables) each one containing some data about all objects of this type existing in the system. One of the key points is the fact that the usage of complex (from the user's viewpoint) refined names of R-variables and their attributes makes it possible to preserve the semantics of complex data structures represented in the form of a set of relation values.

## Part 2. R*O system from inside. Possible realization.

The usage of *n*-ary relations as a only admissible data structure is the basis of the formal relational data model. However, this structure is insufficiently expressive for creation of an adequate model of an enterprise. The approach proposed in the first part resolves this contradiction by allowing us to represent data described as a set of identifiable interconnected complex objects of inherited polymorphic types in the form of a set of relations.

In the second part, we are going to show that the R*O system can be created on the basis of existing relational DBMS. Thus, ORM claims that the relational database management systems may evolve in the direction determined, first of all, by the necessity of adequate description of complex enterprises and suggests a way of this evolution. This assertion is based on the approach described in the first part. This approach deals with a *dual* data representation where the same data are simultaneously represented as values of objects and as values of R-variables. Generally speaking, the implementation of the system proposed below is based on the assumption that, since R-variables are nothing else but relation variables, as a basis, we can use a system, in which these variables are in a way implemented.

Remark. Considering the principal possibility of the proposed implementation, we do not consider the problems of its performance and efficiency.

Speaking about existing RDBMS, first of all, ORM means the SQL BDMSs. ORM understands that some properties of the SQL DBMS do not agree with the relational model of data [M3, 5]. However, from our point of view, this fact is unimportant from the applied standpoint. We proceed from the fact that the these SQL-DBMSs …
- somehow or other allow ones to implement the main features of the relational model of data despite a certain disagreement with it. This means that there commands exist that make it possible to manipulate with the data represented in a set of relational variables (or, in terms of SQL, tables or view), in particular, to determine the scheme of these variables (table or view headers) and data integrity constraints (keys) and to perform operations defined in the relational model of data over the values stored in these variables;
- are systems of long-term data storage. This means that there exist commands, which create and delete tables intended for long-term data storage and manipulate with data stored in these tables. The system has a set of predefined tables intended for storing metadata, which describe the stored data.

- Are programmed systems. This is to say that there exist commands, which allow to define, to store and to execute of a sequence of commands. Here, we actually mean procedural extensions of SQL;

Note that some possibilities of the implemented R*O system are completely determined by the possibilities of the RDBMS being used. In our opinion, the most important property is the property of consistency of the stored data.

As has been said in the first part of the study, data in the implemented R*O system are represented by values of object components and, simultaneously, by values of R-variables. We will refer to the realized totality of objects and R-variables as the data <u>representation level</u>. It should be understood that the data representation level is realizable and, hence, is virtual. One may speak about the existence of objects and R-variables of the representation level only as far as there exist a collection of commands for manipulation with the objects and R-variables (including the valued stored in them) and a program executing these commands. Receiving a command, this program transforms (translates) it into a command or a sequence of commands of the RDBMS being used. Executing these commands, the RDBMS manipulates with the data stored in the tables. The set of relation variables realized by the RDBMS (i.e., tables) will be referred to as the data <u>storage level</u>. Note that the data at the storage level are nothing else but a relational database.

Thus, the most important part of the R*O system is a translator that transforms commands of the R*O system into commands of the RDBMS being used.

<u>Remark.</u> Demonstrating the possibility of translation of commands of the representation level into commands of the RDBMS, in one way or another we mean the possibility of recording the translation results by using SQL. However, this is not necessary from the applied point of view. It is assumed that the translator can generate an internal representation of commands or sequences of commands for a particular RDBMS without creating an SQL script.

It is clear that the system must check the object structures during their whole life in such a way that this structure does not contradict the object description. Executing a command that manipulates with an object (i.e., translating it), the system must have information about the structure of this object. Therefore,
(1) the system must keep the description of the structure of object types during all the time of their existence (a possible scheme of directory tables used for this purpose is proposed below);
(2) it is necessary that, for any object, it is possible to determine the type of this object. A realization of operations **o IS t** and **o OF t**, which fulfill this requirement, will be considered below.

Repeat that the main idea, on which the proposed realization of R*O system is based, is that the R-variables (the data representation level) are relation variables. However, it is important to understand that the R-variables of the representation level and the relation variables of the storage level are not equivalent. Detailed data organization at the storage level is hidden from the user. In particular, the value of R-variable of a component of type **t.a** (representation level) is realized, in the general case, by the system as the joint of several values, one of which is the value of the basic variable $_{st}$**t.a'** (this variable is used for storing the values of the object components which are implemented at the representation level as stored).

<u>Remark.</u> The scheme of the basic variable of storage level $_{st}$**t.a'** completely coincides with the scheme of R-variable of representation level and is the scheme of the corresponding component extended with the OID attribute.

The other values may be treated as an intermediate result of the expression **f'(BASE_VALUE)** corresponding to the object components, which are implemented at representation level as calculated

**t.a = $_{st}$t.a' UNION $_{st}$f'(BASE_VALUE) UNION $_{st}$f'$_2$(BASE_VALUE) UNION …**

Here, $_{st}$**f'(BASE_VALUE)** is an expression calculating a value by using as parameter (in the general case) the value (the state) of the database at the storage level. As has been said, the system must perform these calculations implicitly for the user on the basis of the information about the type inheritance and about the implementation of components of these types.

**Translation of expressions**

It is an interesting question where the calculating expression $_{st}$**f'** is taken from and what it is. Without doubts, the expression $_{st}$**f'** is determined by the calculating expression **f** being defined at the representation level; i.e., the matter in question is the translation of the expression **f** defined at the representation level into the expression $_{st}$**f'** executed at the storage level.

<u>Remark.</u> The value of the operand $_{st}$**t.a'** may be treated as the result of the simplest operation which returns the value of the variable $_{st}$**t.a'**. This operation is defined by the expression **AS STORED** specified at the representation level. It shows that the component is implemented as stored.

What is the translation? Emphasize that the expression **f** is a sequence of operations, which may use, as operands, components of a certain, quite *concrete*, object of type **t**. On the other hand, on the basis of properties of the RDBMS being used, the expression $_{st}$**f'** is a sequence of operations defined in the relational data model, which use, as operands, relation variables of the storage level each one containing some information about the set of objects.

Consider conventional OO programming languages. Describing a method manipulating with object attributes, we mean that this method will be invoked for a some object. For instance, in C++, this concrete object is defined in the method body by the optional key word **this** (in essence, **this** is nothing else but the reference to this concrete object). Translating this method, we obtain a procedure in the machine language. A required parameter of this procedure is the value representing the address of an interval of the machine memory; we assume that this interval of the memory stores the data of a this concrete object. In the first approximation, R*O translator's work is based on a similar principle. This means that, sending the object method at the input of the translator, at its output, we obtain a procedure receiving the value of OID as a required argument.

But, in our opinion, of significant interest is the fact that the used RDBMS realizes the set operations determined by the relational model. On this basis, one can show that an expression **f** defined at the representation level can be translated into an expression $_{st}$**f'** of the storage level such that its single execution (no iterators!) results in the changes of the system such as if the initial expression **f** were executed for every object of type **t** existing in the system.

This idea can be illustrated by the following simple example. Suppose that, in the body of method **method** of type **t**, the value of a stored component $a_1$ is assigned to another stored component $a_2$ (we assume that these components are of the same type)

**ALTER CLASS t REALIZE method(…)… AS**
**BEGIN**
    …;
**this.$a_2$ := this.$a_1$;**
…;
**END**

Consider this operation in terms of R-variables **t.a₁** and **t.a₂**. After its execution, the variable **t.a₂** contains tuples whose values are exactly equal to the selection result from relation **t.a₁** by the value of reference **this**. Therefore, the procedure $_{st}$**method'** of the storage level corresponding to this method must take approximately the following form:

**CREATE PROCEDURE $_{st}$method'(this' …, …)…**
**BEGIN**
   …;
   **INSERT INTO $_{st}$t.a'₂ VALUE (SELECT * FROM $_{st}$t.a'₁ JOIN this');**
   …;
**END**

It is assumed that the value of parameter **this'** intended for transmitting the OID of the object, in which this action is executed, into the procedure is a unary relation with the unique attribute OID.

There are no doubts that the parameter **this'** here may contain OID of a set of objects; in this case, the described action will be *simultaneously* executed for all these objects. It is this fact that allows the group invoke of the method. For instance, when invoking
**t<cond>.method(…)**
the action described will be executed for all objects of type **t**, which satisfy condition **cond**.

This case is the simplest one. Consider the principles of translation in the general form.

**Propositions on compilability**

ORM claims that any relational operation **f** specified at the representation level (where **f** is a superposition of set-theoretic and special operators of relational algebra [3, 4]) on components **a** of an object **o** of type **t** can be transformed into an operation $_{st}$**f** such that, applying it to the values of basic variables of the data storage level, we obtain the relation value, which is the union result of the application of the operation **f** to all objects of this type existing in the system (proposition on R-compilability).

**For any operation f, res = f(o.a₁ ,…, o.aₙ), there exists an operation $_{st}$f, res' = $_{st}$f'($_{st}$t.a'₁, …, $_{st}$t.a'ₙ) such that res = (res' WHERE OID = o)[!OID]**

First, we prove that, for any operation **f, res = f(o.a₁ ,…, o.aₙ)**, there exists an operation **f', res' = f'(t.a₁, …,t.aₙ)**, such that **res = (res' WHERE OID = o)[!OID]**. In other words, we show that the operation **f** can be transformed into an operation **f'**, which uses, as operands, the values of R-variables of components of type **t.a** of the data representation level (the operands of the operation $_{st}$**f** are the values of the basic variables $_{st}$**t.a'** of the storage level).

Recall that the value of **o.a** is calculated as **(t.a WHERE OID = o) [!OID]**. It can be shown that, for the following primitive operations, the following assertions hold:
(1) the union (operands are scheme consistent)   **o.aᵢ  UNION o.aⱼ**
   is equivalent to **( ( t.aᵢ UNION t.aⱼ) WHERE OID = o ) [!OID]**
(2) subtraction (operands are scheme consistent) **o.aᵢ  MINUS o.aⱼ**
   is equivalent to **(( t.aᵢ MINUS t.aⱼ) WHERE OID = o ) [!OID]**
(3) the Cartesian product **o.aᵢ  TIMES  o.aⱼ**
   is equivalent to **( (t.aᵢ JOIN$_{OID}$ t.aⱼ) WHERE OID = o ) [!OID]**
   (here, **JOIN$_{OID}$** is joining by the **OID** attribute)
4) selection **o.aᵢ WHERE** *condition*
   is equivalent to **( (t.aᵢ WHERE** *condition***) WHERE OID = o) [!OID]**

(here, *condition* denotes a condition)
5) projection $o.a_i[r_{a1}, …, r_{an}]$
   is equivalent to $(\,(t.a_i[r_{a1}, …, r_{an}])$ WHERE OID = o$)\,$[!OID]

Thus, the condition being proved holds if **f** is one of the primitive operations listed above. However, since the relational algebra is closed (i.e., the results of an operation can be an operand of another operation) and any operator can be represented as a complex superposition of primitive operations, it can be shown by induction that the proved assertion holds for any relational operations.

Now, we show that the operation **f'**, which uses the values of R-variables of components of type **t.a** of the data representation level as operands, can be transformed into an operation $_{st}$**f'** whose operands are the values of basic variables (tables) $_{st}$**t.a'** at the storage level. We are based on the fact that a value of an R-variable is implemented in the system as the union of several values, one of which is a value of the basic variable $_{st}$**t.a'** of the storage level used for storing the values of the object components, which are implemented at the storage level as stored. The other values may be treated as an intermediate result of the expression $_{st}$**f'(BASE_VALUE)**, corresponding to the object components, which are implemented at the representation level as calculated

**t.a** = $_{st}$**t.a'** UNION $_{st}$**f'**$_1$(…) UNION …

here, $_{st}$**f'**$_i$(…) is nothing else but the R-translation of the operation **f**$_i$ defined at the data representation level. Assuming that this operation uses the values of components of the object as operands **f**$_i$(…, **o.a**, …), we may claim that, in accordance with the first part of the proof, we have

**t.a** = $_{st}$**t.a'** UNION **f'**$_1$(**t.a**$_1$, …, **t.a**$_n$) UNION …

Therefore, again by induction, one can show that the value of any R-variable **t.a** can be actually calculated via the values of basic relation variables of the data storage level.

**t.a** = $_{st}$**t.a'** UNION $_{st}$**f'**$_1$($_{st}$**t.a'**$_1$ UNION **f'**$_{11}$($_{st}$**t.a'**$_{11}$, …) UNION …) UNION …

**Corollaries of the proposition on compilability**

<u>Preliminary remark 1.</u> As has already been said, a change of the state of objects consists in a change of values of their components. A change of the component value means that a new value is assigned to this component (the values themselves are unchangeable). The value of component **a** of object **o** is changed by the assignment operation
**o.a := rval**
where **rval** is the value of a relation whose scheme coincides with the scheme specified for the component **a**.

Note that the operation conventional for the existing relational DBMS, which change values of the relation variables (insertion and deletion of tuples, as well as the update operation changing the attribute values of tuples existing in the variable), can be written via the assignment operation:
- the insertion operation **INSERT o.a$_i$ VALUE(rval)** may be represented as **o.a$_i$ := o.a$_i$ UNION rval**,
- the operation of tuple deletion **DEL o.a$_i$ WHERE (condition)** may be represented as **o.a$_i$ := o.a$_i$ WHERE (NOT condition)**,

- the operation of changing the attribute values **UPDATE o.a$_i$ SET x$_j$=s WHERE(condition)** may be represented as **o.a$_i$ := (o.a$_i$ WHERE (NOT condition)) UNION (o.a$_i$ WHERE (condition)[…, x$_{j-1}$, s, x$_{j+1}$,…]).**

Conversely, the assignment operator
**relvar:=relvalue**
can be implemented via the operations conventional for the existing relational DBMS
**DELETE FROM relvar;**
**INSERT INTO relvar VALUE relvalue;**

Preliminary remark 2. As has already been said, methods of classes can contain local variables of value types. Consider a possible implementation of these variables in the BDMS being used.

Local variables serve to store the values used in a method and/or appearing while executing the method. If the actions described by a method **method()** are executed simultaneously for a set of objects (no iterators), then we may speak about the set of such values and, accordingly, about a set of variables **localvar** intended for storing them; moreover, one variable and, at each time instant, one value of this variable correspond to each object of this set.

Therefore, we may say that the local variable **localvar** with the scheme **(…, x$_k$:D$_k$, …)** at the storage level must be associated with only one relation variable $_{st}$**localvat'** with the scheme **(OID: DOID , …, x$_k$:D$_k$,…)**. Thus, the implementation of local variables at the storage level does not differ from that of stored components. The only difference between them is that the variable of the storage level $_{st}$**localvar'** corresponding to the local variable is a temporary variable whose lifetime is limited by the execution time of the R-translated procedure $_{st}$**method'()**. This allows us to claim that the following corollaries hold both for components and for local variables of methods.

Corollary 1 (on R-compilability of the assignment operation). Suppose that the value of component **a$_k$** of an object of type **t** specified by reference **o** is the result of an operation **f** on other components **a$_i$** of this object. In other words, we deal with the assignment operator changing the state of component **a$_k$**
**o.a$_k$ := f(o.a$_1$,… , o.a$_n$, …).**  (1)
Then, by virtue of the proposition on R-compilability of operations, this operator is associated with the operator
**t.a$_k$ := f'(… , t.a$_i$, …),** (1')
which changes the value of the R-variable **t.a$_k$** corresponding to the component **a$_k$** as operator (1) were executed for every object of type **t** existing in the system. As has been said, in the existing RDBMS, such an operator can be implemented by the pair **DELETE …, INSERT …**

Corollary 2 (on R-compilability of a sequence of the assignment operations).

Suppose that, for an object of type **t** specified by reference **o,** a simple sequence of operators of form (1) is defined
**o.a$_j$ := f$_1$(… , o.a$_i$, …);**
**o.a$_k$ := f$_2$(… , o.a$_j$, …);**.
**o.a$_l$ := f$_3$(… , o.a$_k$, …);**

In the R-representation, this sequence of operators is associated with an absolutely analogous sequence of R-translated operators of form (1')
**t.a$_j$ := f$_1$(… , t.a$_i$, …);**.
**t.a$_k$ := f$_2$(… , t.a$_j$, …);**.

$t.a_l := f_3(\ldots, t.a_k, \ldots);$

which changes the state of the system in such a way as if sequence (2) were executed for every object of type **t** existing in the system.

ORM asserts that any sequence of operations on components **a** of an object of type **t** specified by a reference **o** can be R-translated into a sequence of operations on corresponding R-variables **t.a** such that its single execution changes the state of the system in such a way as if the original sequence were executed for every object of this type existing in the system. As examples confirming this assertion, consider the R-translation of the conditional operator and the loop operator.

Note that any operation $f_i$ returns a value $p_i$, which, being further used, must be stored in an appropriate local variable (maybe, temporary and/or not defined explicitly). Suppose that the operation returns a value of the Boolean type and an appropriate variable **b** is used for storing this value. At the storage level, this variable is associated with a relation variable $_{st}$**t.b'** with the scheme **(OID, b)**.

**Conditional operator**. Suppose that, for an object of type **t** specified by a reference **o**, the following sequence of operations is defined:

**IF $f_1(\ldots, o.a_i, \ldots)$ THEN $o.a_k := f_2(\ldots, o.a_j, \ldots)$;**

We assume that the result of execution of $f_1$ is a value of type BOOLEAN. This sequence may be rewritten as

**b := $f_1(\ldots, o.a_i, \ldots)$;**
**IF b THEN $o.a_k := f_2(\ldots, o.a_j, \ldots)$;**

that can be R-translated, for instance, into the following sequence of actions on R-variables:

**t.b := $f'_1(\ldots, t.a_i, \ldots)$;**
**$t.a_k :=$   ( $f'_2(\ldots, t.a_j, \ldots)$ JOIN$_{OID}$ ($_{st}$t.b' WHERE b = TRUE )) UNION**
              **( $t.a_k$ JOIN$_{OID}$ ($_{st}$t.b' WHERE b = FALSE));**

**Loop operator**. Suppose that, for an object of type **t** specified by a reference **o**, the following sequence of operations is defined:

**DO …**
     **$o.a_k := f_2(\ldots, o.a_j, \ldots)$;**
     **…**
**WHILE $f_1(\ldots, o.a_i, \ldots)$;**

Just as in the previous case, the result of execution of $f_1$ is a value of type BOOLEAN. This sequence may also be rewritten as

**DO …**
     **$o.a_k := f_2(\ldots, o.a_j, \ldots)$;**
     **…**
     **b := $f_1(\ldots, o.a_i, \ldots)$;**
**WHILE b;**

that can be R-translated, for instance, into the following sequence of actions on R-variables:

```
DO  …
    t.aₖ :=  ( f'₂(… , p.aⱼ, …) JOIN_OID (ₛₜt.b' WHERE b = TRUE )) UNION
             ( t.aₖ JOIN_OID (t.b WHERE b = FALSE));
    …
    ₛₜt.b' := f'₁(… , t.aᵢ, …);
WHILE EXIST ₛₜt.b' WHERE b = TRUE;
```

Thus, for operations represented by sequences of actions on components of an object of type **t**, the group execution of the operation for all objects of this type is possible, for example,

```
EXECUTE t.somemethod();
```

We assume that such an expression must be translated into a single execution of the R-translated sequence `somemethod'`.

<u>Corollary 3 (on R-compilability of a sequence of the assignment operations for an explicitly defined group of objects).</u>

It can be shown that all propositions about compilability of relational operations and sequences of such operations may also be applied to an explicitly specified group of objects. We suppose that the group of objects is specified by a group reference **g**. Indeed, any R-translated operation

**t.aⱼ := f'(… , t.aᵢ, …)**

may be treated as a particular case of the operation

**t.aⱼ := (f'(… , t.aᵢ, …) JOIN g)  UNION (t.aⱼ JOIN (objects(t) MINUS g)),**

where **g** contains references to all objects of type **t** existing in the system, т.е. **g = objects('t')**. Note that, if **g** defines an improper subset of objects of type **t** existing in the system, then a single execution of the last operation changes the state of the system in such a way as if the initial operation **f(… , o.aᵢ, …)** were executed for all objects belonging to this subset.

Therefore, we may assert that any sequence of operations on components **a** of an object of type **t** specified by a reference **o** can be R-translated into a sequence of operations on the corresponding R-variables **t.a**, which changes the state of the system in such a way as if the initial sequence were executed for every object of the set defined by the group reference **g**. As has been said, this set is a proper subset of all objects of type **t** existing in the system.

Thus, there is a possibility of a group invoice of operations which can be represented as a sequence of actions on components of an object of type **t**, for the whole set of objects specified by the group reference **g**, for instance,

```
EXECUTE g.somemethod();
```

or for a group determined, for instance, by the reference `xref` from the selection of objects of type **t** by value

```
EXECUTE t<cond1,...>.xref.somemethod();.
```

**Catalogue**

Describing principles of the catalogue organization, ORM assumes that the used RDBMS contains a catalogue that allows one to describe the scheme of its "table" memory (i.e., the scheme of the relational database). This scheme contains a description of relation variables at the storage level whose structures are defined and uniquely correspond to the component schemes specified and described at the representation level. Therefore, we will not consider in detail the catalogue parts containing some information about the structure of these relations and restrict our consideration by the assertion that, with each such variable, an identifier $_{STORED}R$ should be associated, which is defined on the domain of relation identifiers $_{STORED}RVarIDs$ (here and below, identifiers may be, for instance, unique names).

In the most general form, a catalogue describing the structure of types must contain the following tables (the fields included in the primary key of the table are underlined)
1) Table **valTYPES(vT:vTypes …)** enumerates the value types existing in the system (**vTypes** is a domain of identifiers of value types, for instance, type names). These types may be described in detail in other tables (not described here).
2) Table of object types **objTYPES (oT:oTypes …)** enumerates all object types existing in the system (**oTypes** is a domain of identifiers of object types, for instance, type names).
3) Table **IS_T(oT:oTypes, IS_oT: oTypes …)** completely describes the inheritance relation existing between the object types. For any type **t**, in the table, there exists a record **oT** = **IS_oT** . Moreover, for any type **t**, in the table, all its direct and indirect basic types **IS_oT** are enumerated.
4) Table **SPEC(A:Comps, oT:oTypes, vT: vTypes, signature …)** enumerates the own (i.e., uninherited) components (field **A**) of object types (field **oT**) and shows the value type of these components (field **vT**) and other information determining their signature (**CompID** is the domain of component identifiers, for instance, component names). Actually, this table contains the complete information about the specification of own components of object types.
5) Table **REAL(A:Comps, OF_oT:oTypes, isSTORED:BOOLEAN, RealExpr…)** contains information about the implementation of components (field **A**) of object types (field **OF_oT**) and, in particular, the information whether an attribute is stored (field **isSTORED = TRUE**) or calculated (field **isSTORED = FALSE**). In the first case, the field **RealExpr** contains the name of the variable of the storage level $_{STORED}R$ used for storing and, in the second case, the translated calculating expression or function. (The **REAL** table may contain the information only about own type components and those redefined while inheriting. For components that are not redefined while inheriting, a mechanism should exist, which returns the information about realization of this component in the nearest basic type, where this realization was explicitly specified.)

Of course, all aforementioned tables may contain other fields containing some information about the specification and realization of object types.

The operations of the data representation level, which manipulate with the data scheme, are translated into operations manipulating with catalogue tables.

The command creating a new object type

```
CREATE CLASS otypename [EXTENDED parenttypename[,parenttypename] ]
{
      signature
      [;signature]
}
```

must be executed as follows.

1) In table **objTYPES**, a single record about the new type is added. It contains the name (**otypename**) of this type.
2) In table **SPEC**, records are added, which contain specifications of the signatures of the type components and methods, whose number is equal to the number of own components and methods defined in the type.
3) In table **ISt**, at least one record is added. One record is added for the type without basic type (for simplicity, we omit here the required inheritance from the dummy type Object). For types that inherit the types existing in the system (**EXTENDED parentclassname**), it is also necessary to enumerate all (direct and indirect) basic types in table **ISt**. Note that the information about indirect basic types can be obtained from records (contained in the same table **ISt**), which describe the direct basic types.

The operation of changing types can add, change, and delete specifications of own attributes and methods of the type (i.e., change the specification),

```
ALTER CLASS otypename ADD| DROP|ALTER signature_name;
```

as well as add, change, and delete *realizations* of own and inherited attributes and methods of the type (in accordance with the existing specificaton).

```
ALTER CLASS otypename REALIZE signature AS realize_expr;
```

These operations are implemented as addition, change, and deletion of records in the tables **SPEC** and **REAL**, respectively.

The operation of deletion of an object type

```
DROP otypename;
```

executes actions converse to actions executed while adding and deletes records about the realization of attributes and methods of this type.

**Table of identifiers**

The table of identifiers **OIDS( OID: tOID: OF_oT:oTypeIDs)** enumerates the unique identifiers of objects existing in the system and, with each object identifier (field **OID**), associates an identifier existing in the catalogue (field **OF_oT**) of the object type of this object.

Note that the **OID** field, which is present in tables of the storage level, as well as any fields defining the existence of a link by reference, must be declared as an external key, which refers to the **OID** field of the table of identifiers. This allows the following:
1) To check the integrity of links "by reference" using for this purpose the mechanisms for checking the reference integrity existing in the used relational database.
2) Claim that the system contains no records of data tables, for which the object identifier is not defined (i.e., any record in the data tables actually describes an object). Note that, in existing RDBMS, which realize the cascade data deletion, the deletion of a record with OID of an object from the identifier table implies the deletion of related records containing data of this object.

In turn, the field **oT** must be declared as a foreign key related to the field **oT** of the catalogue table **objTYPES**. This guarantees that, for any object in the system, the type of this object is defined. The existence of this link allows one to use the mechanisms existing in the RDBMS being used, which allow one, by executing the **DROP otypename** command, to delete all objects of this type existing in the system.

The operator **o OF t** described above may be implemented as

**EXIST OIDS (WHERE OF_oT = t AND OID = o)**.

To implement the **o IS t** operator, it is necessary to use the information about the type inheritance (see description of the catalogue table **IS_T**)

**EXIST (OIDS JOIN$_{OF\_oT = oT}$ IS_T(WHERE OID = o AND IS_oT = t))**

**Implementation of control commands**

The totality of commands intended for controlling the R*O system is a nonprocedural high level language, which should be considered as the main (maybe, unique) way tool for access to the data stored in the system. The commands of this language can be partitioned into two groups, one of which is the data declaration sublanguage (DDL) and the other is the data manipulation sublanguage. We assume that the commands of this language can be translated into commands of the RDBMS being used.

We have considered the actions that should be performed by the main DDL commands as operations on the system catalogue. As has already been said, nonprocedural commands determining the realization of components or flip-flops contain calculating expressions. These procedural expressions are translated into procedures of the RDBMS being used, which receive, as a required parameter, the object identifiers determining an object or a group of objects. These procedures are stored in the catalogue table. From this table, they can be loaded for execution. Thus, when processing the procedural extensions of DDL, the translator plays the role of a compiler that transforms R*O code into the code of the RDBMS being used.

The commands of the data manipulation sublanguage (DML) are intended for creation and deletion of data objects, as well as for controlling the state of these objects and for obtaining the information about the data stored in the system. Note that, processing and executing DML commands, the translator plays the role of an interpreter.

The command `NEW t(constructor_parameters)` intended for creating new objects of type **t**. Receiving this command, the system generates a new OID and writes it together with an identifier of type **t** into the identifier table **OIDS**. Then, on the basis of the description of the object structure in the catalogue of types, the system adds the tuples intended for stored components of the new object to the basic relations of the storage level. Here, the **OID** attribute of these tuples is initialized by the object identifier of the created object. Then, if necessary, the constructor is invoked.

Receiving the command `DESTROY objectgroup`, the system must fulfill actions converse for those executed while processing the `NEW` command. Recall that the system may control the reference integrity (see the chapter *Table of identifiers*); i.e, in particular, an object referenced from the other parts of the system cannot be deleted.

The state of objects may be changed by directly changing the values of components of these objects
```
INSERT … INTO objectgroup.a;
UPDATE objectgroup.a;
DELETE FROM objectgroup.a;
```

or by invoking some methods,
```
EXECUTE objectgroup.methodname(parameters);
```

To retrieve data stored in the system, we use group data access commands applied to R-variables. These commands are based on
- known operations of relational algebra,
- OV-retrieval operations and operation of reference expansion,
- group method invokes, and
- superposition of all operations listed above.

The translation of commands changing the object state is determined by the proposition on translation and its corollaries.

## Conclusions

The OverRelational Manifesto confirms the most important positions of its predecessors. Similarly to "The Object-Oriented Database System Manifesto" it supports the idea of long-term stored complex objects. Similarly to "Third-Generation Data Base System Manifesto" it assumes that the existing systems for data storage can be used and developed. Moreover, similarly to "The Third Manifest," it tends to preserve the purity of ideas of the relational data model.

On the basis of the approach proposed by ORM, a system can be developed, which can be treated, first of all, as a system for creating an adequate, active, and long-term model of the enterprise that is controlled by the user and provides the user with data about its state.

The author hopes that this study may be useful for specialists and programmers working at the joining point of studii related to data storage systems and data modeling. Of course, bounded by the paper size, we cannot present the theme in more details: some questions are only outlined, some other are omitted. Without doubts, many questions related to this approach are even not posed. However, the fact that the approach proposed is based on the relational data model with formal mathematical foundation allows us to hope that comprehensible answers to these questions may be obtained in formal way.